\documentclass[10pt,conference]{IEEEtran}
\usepackage{eurosym}
\usepackage{amssymb}
\usepackage{amsmath}
 \usepackage{paralist}
\usepackage{hyperref}
\usepackage{graphicx}
\usepackage [noadjust]{cite}
\usepackage{color}

\begin{document}
\title{\textbf{\Large Data Validation for Big Live Data}}
\author{
\IEEEauthorblockN{Malcolm Crowe, Carolyn Begg}
\IEEEauthorblockA{School of Computing\\
University of the West of Scotland\\
Paisley PA1 2BE, UK\\
email:\,\{malcolm.crowe$\mid$carolyn.begg\}@uws.ac.uk}
\and
\IEEEauthorblockN{Fritz Laux}
\IEEEauthorblockA{Fakult\"{a}t Informatik\\
Reutlingen University\\
D-72762 Reutlingen, Germany\\
email: fritz.laux@fh-reutlingen.de}
\and
\IEEEauthorblockN{Martti Laiho}
\IEEEauthorblockA{DBTechNet\\
www.dbtechnet.org\\
email: martti.laiho@gmail.com}
}

\maketitle
\begin{abstract}
Data Integration of heterogeneous data sources relies either on periodically transferring large amounts of data to a physical Data Warehouse or retrieving data from the sources on request only. 
The latter results in the creation of what is referred to as a virtual Data Warehouse, which is preferable when the use of the latest data is paramount. 
However, the downside is that it adds network traffic and suffers from performance degradation when the amount of data is high. 
In this paper, we propose the use of a readCheck validator to ensure the timeliness of the queried data and reduced data traffic. 
It is further shown that the readCheck allows transactions to update data in the data sources obeying full Atomicity, Consistency, Isolation, and Durability (ACID) properties. 
\end{abstract}

\IEEEpeerreviewmaketitle

\section{Introduction}
\label{sec:intro}

\begin{figure}[]
\centering
\includegraphics[width=0.40\textwidth]{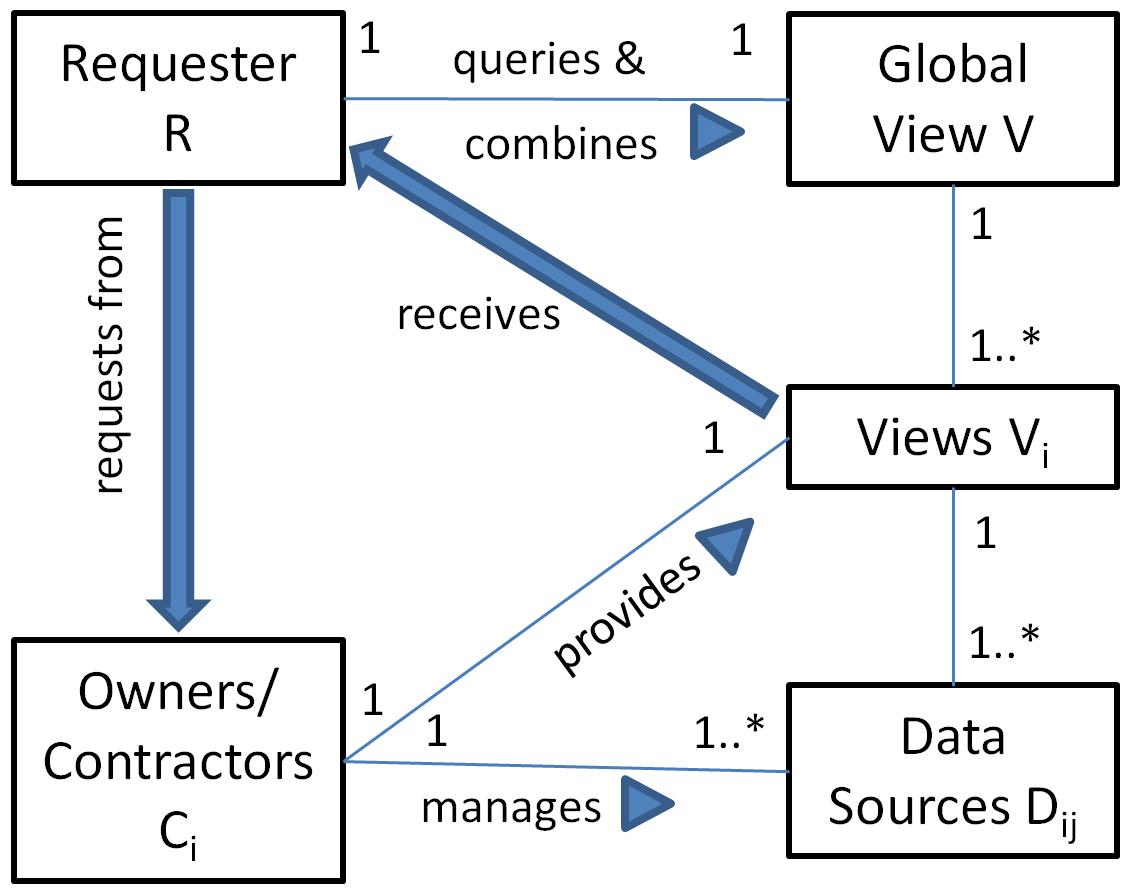}
\caption{Schema and relationships of the virtual data integration scenario}
\label{fig:SchemaModel}
\end{figure}

For the Data Integration scenario (see Figure \ref{fig:SchemaModel}), we assume a set of heterogeneous data sources \{$D_{ij}$\} belonging to and managed by a (disparate) set of contracting parties \{$C_i$\}, which  provide Views $V_i$ for a Requester $R$ (e.g., a regulatory body, enterprise, or government). 
The  databases that store the data in a variety of formats adopted by the different contractors, remain under the control of their respective owners. 
We also note that in our example the regulatory body $R$ is normally concerned with aggregated data rather than with individual records and the contractors are generally also responsible for the privacy and security of the data they hold. 
But, importantly, in all our examples $R$ is concerned with the current situation and up-to-date aggregations are required and queried from a Global View $V$ of the \emph{live} data set \{$D_{ij}$\}. 
These requirements make it undesirable to create and store a single big data set at $R$.

To illustrate the situation let's take the Ebola outbreak in West Africa in 2014. The World Health Organization (WHO) takes the role of $R$ in our model and $C_1$ is given by the Choithram Memorial Hospital in Freetown.
The demographic information about Sierra Leone shall be provided by Statistics Sierra Leone, $C_2$ in our model. There are more hospitals $C_i$ in Sierra Leone, Guinea, and Liberia as well as official statistics offices. We leave them out to not overload our example.
$C_1$ records all patient data, diagnosis and treatment but provides only non-sensitive data as an aggregated view $V_1$ to the WHO. $C_2$ also provide a view $V_2$ of a portion of the statistical data collected. To not complicate the integration we assume that both views provide the residence of the patient and the location (city/quarter or village) in the same coding.

With $V_1$ and $V_2$ provided, $R$ can build an integrated view $V$ providing data that allow the analysis of the distribution and spreading of the disease.
It is evident that the data must be up-to-date (live data) in order to monitor the spreading and allow decisions for improving quarantine and treatment.
We continue this example in more detail and implement it in Section \ref{ssec:globSchemaD}.

Big Live Data as discussed here consists of data sets that are subject to real-time updates (\emph{live} data) and \emph{big} not just in terms of size in bytes but also in the sense that they span multiple areas of responsibility and ownership. 
The scenarios we have in mind include international and other regulatory bodies, where data belong to and is managed by separate entities (including governments), and in business examples such as supply chain, manufacturing, project management and construction, where each contractor or company produces their data but an overseeing body must have access to a global view of this data to support decision-making at the highest level to enable the management and coordination of the data contributors.

The term "Big Data" is usually applied to scenarios characterised by the "4 V’s". We consider that all of these apply to the scenarios presented here:
\begin{itemize}
 \item \textbf{Velocity} – refers to the significant volumes of new data that can be created and/or updates made at speed in the individual $D_{ij}$. Because of velocity, it is not feasible to keep making fresh collections of all data at $R$.
 \item \textbf{Volume} – refers to the significant volumes of data held across many separate data sources $D_{ij}$, which are required wholly or in part to provide the \emph{live} data for $R$. 
 \item \textbf{Variety} – refers to the range of possible data models (e.g., relational, NoSQL, XML) and data types  used by the $D_{ij}$. A \emph{live} (on-line access to $D_{ij}$) implementation of $V$ will need to resolve transformation and semantic issues associated with a variety of data types. It is the responsibility of $C_i$ to provide a consistent View $V_i$ and it is $R$'s task to transform these views into a Global View $V$.  
 \item \textbf{Veracity} – refers to the range in the data quality (from high to lower levels) of the data held in the separate $D_{ij}$.  The $C_i$ have a responsibility to ensure the reliability and veracity of their data. A
 \emph{live} implementation of $V$ should aim for maximum coherency and consistency of the overall data collected, but as a minimum, any data at $R$ should be on consistent results from the $D_{ij}$ (resp. $V_i$) as at a particular time.
 \end{itemize} 
In this paper, we refer to the veracity property as correctness, and apply this concept both for current results, which should be consistent and up to date, and for stored results, which should be a correct snapshot of the state of affairs at the time they were computed. 
We note that most Big Data implementations have difficulty with correctness if the data is subject to change.
\subsection{Contribution}
In our scenario, $R$ should also be able to check that the most recent results obtained from the $C_i$ are still correct, and perform other checks for validation, maybe including supplementary data gathered from other sources. 
Greater care over data curation, ownership and provenance in such complex scenarios can help to achieve a higher fidelity of data. 
The present paper proposes the use of a readCheck validator in order to check the timeliness of a query result and reduce the data traffic between the sources and the requester. 
The same mechanism can be used to implement an optimistic concurrency control mechanism for heterogeneous and geographically distributed data sources. 
This enables distributed \emph{live} updates to the underlying data sources obeying full Atomicity, Consistency, Isolation, and Durability (ACID) properties.
\subsection{Structure of the Paper}
With the following overview of Related Work on data integration the context for our validation concept will be settled. 
Section \ref{sec:ConceptDV} introduces the concepts of Row Version Validation (RVV) and ETags for Web data caching, which are the basis for a generalised data validation. 
Then, these concepts are combined and applied to query processing for joined and aggregated data over a virtual Data Warehouse (DWH). 
The technical realisation of the validation mechanism via Representational State Transfer (REST) services is explained in Section \ref{sec:Implementation} and applied to general query processing over a virtual DWH. 
The extended syntax for the global schema design is presented in Subsection \ref{ssec:globSchemaD} and illustrated with an example.
The paper ends with a summary of our findings from the pilot implementation and gives an outlook on ideas for future work.

\section{Related Work}
Since the beginning of 1980 many papers on data integration have been published. 
The research of Inmon \cite{Inmon} and Jarke \cite{Jarke} concentrated on the DWH 
approach using Extract-Transform-Load (ETL) techniques \cite{ Bouzeghoub, Vassiliadis, Simitsis, Karakasidis}, 
 schema matching \cite{Mannino} and integration \cite{Put, Navathe}. 
Kimbal et al. \cite{Kimbal} present methods on how to support the whole DWH live-cycle including schema design, ETL methods, and how to implement and deploy such a system.
Later, the focus changed to real-time ETL or virtual DWH \cite{Srinivasan, Gorawski}. 
Myronovych and Boreisha \cite{Myronovych} discuss how XML Web Services can be used for the ETL process in the context of Enterprise Information Integration. In their textbook Doan, Halevy, and Ives \cite{Doan} cover comprehensively the different concepts of data integration using conjunctive queries as formal representation.

Since for legal and practical reasons, the use of ETL techniques are not available in our case, the focus will be on how $R$ can manage to collect correct information using the REST architectural style proposed by Fielding \cite{FieldingPhD}. 
This will lead us to consider the use of HyperText Transfer Protocol (HTTP) \cite{FieldingHTTP} techniques such as ETags \cite{FieldingRFC7232}, and how these can relate to database transaction concepts. 

In this paper, we turn to the question of data validation. 
If $R$ has collected the result of a query from the set of contractors, and wishes to publish a report, or take action as a result, how can $R$ check that the data obtained from the contractors are still current? 

For performance reasons data is often cached. Usually cached data is only valid during a time defined by a Time-To-Live (TTL) indicator. After that it is considered stale and will be evicted from the cache.  Many works on cache management has been published over the years (DBLP reports 477 matches to "cache management"), but most of the propositions are specific to certain architectures (e.g. OLAP \cite{Marques}, J2EE \cite{Perez-Sorrosal}), require support from the network nodes \cite{Bilal} or are optimised for special use cases \cite{Viana-Ferreira}. Some require column stores \cite{Mueller}, a middle tier \cite{Luo}, \cite{Bernstein} or database support \cite{Ghandeharizadeh} \cite{Asad}. As we do not expect any specific architecture or technology from the data sources, most of these sophisticated caching is not usable for us. 

Caching of query results is however desirable to avoid unnecessary network and processing load.
This is dealt with today on the Internet by considering validators for cached data.
This departure from stateless HTTP is extremely useful in our context because it enables us to set up mechanisms similar to ACID transactions for the extreme cases of distributed data considered here. 
HTTP offers ETags \cite{FieldingRFC7232} in response to allow caching of results, and ETags can be used for validating a step in a transaction. 
ETags are very similar to the RVV concept \cite{Laiho} or Multi-Version Concurrency Control (MVCC), successfully used in PostgreSQL \cite{Kapila}, SQL server \cite{Howard}, Oracle  etc. to provide optimistic execution. 
In fact the ETag can coincide with the RVV validator for requests that return only one row of a base table.

Xiong Fengguang and his colleagues \cite{Fengguang} present a framework for virtual data integration of heterogeneous data sources using XML as interface. Jinqun Wu \cite{Wu} implements the approach using Web services as data adapters. The adapter also provides access to metadata which is helpful for data discovery and query optimization. The implementation uses two caches, one for the metadata and the other for query parsing and result caching. However, it is not explained, how the freshness of data is ensured. As XML tends to be very verbose the performance of the conversion to and from XML is unclear.
Naoki Take et al. \cite{Take} also propose virtual integration for operation support systems. The authors argue to use a mediated relational schema as integration basis. A wrapper maps event data to a virtual table which can be accessed if the necessary parameter is provided in the WHERE-clause. This is similar to our approach, but we prefer the REST style for accessing non-relational resources.
The performance results confirm that queries to a virtual integration database are one magnitude slower than to a materialized one. This clearly calls for some caching when using a virtual mediated schema. 

\section{Concepts for Data Validation}
\label{sec:ConceptDV}
First, we present the ideas of RVV and ETags as basis for our general validation mechanism,  called \emph{readCheck}. Second, the readCheck is used to validate the freshness and consistency of queried data and build an optimistic concurrency control protocol on it. 

\subsection{Row-version validation}
\label{ssec:RVV}
The RVV protocol is a type of version control mechanism, which can be used for a form of optimistic concurrency control, alongside or in preference to other versioning measures such as MVCC.
The model implementations of RVV in the Laiho/Laux paper \cite{Laiho} envisage a sequence generator to ensure uniqueness of RVV values, so that new values of this sequence are added as a special row-version column in base tables on each INSERT or UPDATE.  
Some Database Management Systems (DBMSs) include row versioning mechanisms that can be used for this. 
Otherwise the code for doing this is implemented as database triggers on these tables. 

Our scenario is a bit different as such a guarantee is hard to provide by the contractors (not all data sources $D_{ij}$ need to be databases). On the other hand as we will see, if ETags are part of the service offered to $R$ by the $C_i$ they can provide a suitable version stamp instead.
In an \emph{RVV transaction}, a read-write transaction or "long transaction" consisting of a sequence of read actions followed by some write actions, we should have the following steps:
\begin{enumerate}
\item The first read action (selection query) also reads and records the version stamp(s).  
Note:  With MVCC or reading from a cache this may already be stale data.
\item The version stamp obtained can be optionally used as a validation predicate in later SQL-operations, or included in a precondition. The precondition would only need to compare the old version stamp with the new one to ensure that the data has not been changed in the meantime. If the version stamp is no longer valid, it is best to start again from the beginning as some of the data being used is already out of date. 
\item In the same way the RVV predicate is then included in the search condition of the write actions (UPDATE or DELETE) to the database (bypassing the cache if any).  
\item If no rows are affected in step 3) the transaction failed. This may be because a) the version stamp obtained in step 1 was already outdated (possible with optimistic concurrency control), b) there have been intervening update(s) by concurrent transaction(s), or c) the row has been deleted by a concurrent transaction.
\end{enumerate}
Case 4c) can be detected by a new read action without the RVV predicate in the same transaction. 
If the result is NOT FOUND then  case 4c) is true, otherwise 4a) or 4b) apply. In these cases, it may be worth starting the RVV transaction again from the beginning with new search values depending on the application.

\subsection{RVV and complex selections}
\label{ssec:RVVcomplex}
RVV is just for one row, and in the Laiho/Laux model is an integer value associated with a base table row that is changed if any change is made to a value in the row. 
If we were to define a view using a join, or embedded arrays, then we could extend the idea of RVV for such a complex row so that it comprises all base table rows selected for that row of the view, and would change if any of these where changed. 
For example in whatever join of two base tables, $J=A\bowtie B$ say, the RVV of each resultant row $j$ of $J$ will include the RVVs from the contributing rows of the base tables.
We could therefore implement RVVs to allow compound values, so that the RVV for a join could be a comma-separated list of values or a vector of integers.
 
If the RVV model of Laiho/Laux \cite{Laiho} is extended in this way it can be used to validate the results of join queries, or more complex selections where a row in the selection embeds values from rows of other tables. 

If the business application wishes to make a change to a derived table (update or delete) such RVVs can then be used
to validate and carry out the operation on the relevant base table rows. 
The semantics of updating a value in a row of a derived table will often be reasonable, and it could be meaningful to support a delete operation on a derived table. A use case for updating or deleting data could arise from new or changed/corrected information that arise from sources external (e.g., a regulatory body or multinational organisation)  to the owner of data source. 

All of the above in our scenario remains within an individual $V_i$, and an RVV as seen from $R$ could be extended to identify the contractor(s) involved, so that the results of R's queries could be updatable, depending on the permissions granted to $R$.

\subsection{ETags as version stamps}
\label{ssec:ETagVS}
As not all data sources are databases or have SQL interfaces, another mechanism is needed to avoid stale data.
Fielding and Reschke propose in RFC 7232 \cite{FieldingRFC7232} a header field in an HTTP request, called \emph{ETag}.
This ETag should serve as a validator for the freshness of data.
An Etag should be returned for any GET request for a Web resource, possibly returning a lot of data. 
If we generalise its use to arbitrary queries, we would like the ETag validator to confirm that the results obtained for the query are still valid. 
In fact, ETags can be used in subsequent requests as preconditions; that is, if a requested ETag is no longer valid, the server should reply according to the HTTP protocol with "412 Precondition failed". 
Thus, the ETag protocol has useful similarities with the RVV protocol above. 
If the GET is a REST request in our scenario, in general the data returned will be related or linked in some way but will not in general come from a single row or base table, but the ETag will be effectively a version stamp for the returned data.
Moreover, if the returned data is for a single row in a base table (or more complex selection as above), then the RVV if available can be used as ETag for the result. 

In our scenario, we consider particular sorts of selection and aggregation queries that combine data gathered from the $V_i$, and if REST is used we can hope to have an ETag $e_i$ from each one that contributes data. This will give $R$ a version stamp $\vec{e}$ for the overall selection or aggregation, by combining the contributing ETags in some suitable way, from which we can extract any required $e_i$ by string manipulation.

Where aggregation occurs it is not really practical or desirable for validators to identify all of the rows that contributed to the result. 
But we would like the ETag to let us discover whether the base tables involved have been modified since our results were computed, by extending it to indicate the extent of information read (e.g., tables, or specific rows if practicable) and for each table the most recent version stamp of the rows accessed. 

For performance reasons the ETag could be applied to a hierarchical data structure, beginning on the detail level of a data element and propagating up to the top level of a database. 
So if a query request is executed and the result has been previously cached,  it is sufficient to ask the underlying data source to test if the ETag of the requested aggregate has changed since the last time. 
If there was no change, the last result is still valid. If the query involves multiple sources and only some have changed, it is possible to build the new result by refreshing only the changed values in the cache.

With this framework, we can specify a Versioned REST protocol for our scenario analogous to the one described in \ref{ssec:RVV}. In the four-step protocol it is only necessary to replace a query with a GET-request and the RVV with the ETag validator. 

\subsection{Management of distributed transactions using readCheck}
\label{ssec:MgntDistribTa}
It is possible to extend ETag and RVV concepts to implement "long transactions" for the virtual integration scenario and call it \emph{readCheck}. 
The readCheck needs to include in addition to the ETag or RVV value a unique transaction identifier (server, database, timestamp, taNo). It is then possible to use the readCheck validator in a similar way as the RVV to provide an optimistic concurrency protocol.
We only need to arrange that readChecks are remembered in intermediate results for any rows that have come from other servers and such validators can be accumulated for checking at commit time.
For simple transactions, where all write actions are delayed to the commit stage of the transaction, the readCheck can be used to guarantee ACID behaviour. We note that some database management systems offer snapshot isolation for transactions, thus effectively delaying all changes to the database to follow the commit process. 

The validation machinery requires the following steps:
\begin{enumerate}
\item readCheck information is accumulated by the contractor for all queries $Q$ that are part of the transaction. 
With proxies (or caching) the values read may already be stale.
\item At any stage, the sequence of queries $Q$ belonging to a transaction can be decomposed to ${Q_i}$ and sent to the respective databases (bypassing proxies or caches) to check that $\vec{r}(Q)$ is still correct. 
If not, the data held by the contractor is stale and the transaction will not be able to commit.
\item  The write action and implied commit needs to be sent to the database itself (bypassing proxies or caches) accompanied by the list of readCheck data. If all readCheck data is still valid the write action is performed. 
The database is only locked while the serialization condition is checked.
\end{enumerate}
It is recommended that the contractor should receive updated readCheck data for the state after the transaction commits.

Assuming all the contractors have a way of rolling back aborted transactions (better than taking a backup beforehand), the readCheck mechanism would support the use of an optimistic two-phase commit (2PC) protocol suitable for distributed transactions. This can be realized as follows:
\begin{itemize}
\item Any serialization conflict will be detected if the readCheck has changed since the start of the transaction. As no changes to the database has be executed, the transaction can simply be aborted.  
\item If no conflict was detected the 2PC is executed. If all participating data sources agree, the write phase is entered otherwise the transaction is aborted (no writes will take place).
\end{itemize}
If during the write phase an error occurs, the affected data source must nevertheless guaranty that the write will be executed after the error is removed. This completes the 2PC protocol. 
  
\section{Implementation of the readCheck Mechanism}
\label{sec:Implementation}
As a slight generalisation of the Laiho/Laux concept of RVV, let us suppose that we have to hand a database implementation in which the transaction serialisation mechanism provides a monotonically increasing integer identifier $r(d)$ that it attaches as the RVV for any affected base table row $d$, and define $r(T)$ as the integer identifying the most recent change to table $T$.

Then, for any query $Q$ on a single database, the readCheck $\vec{r}(Q)$ is a list or vector of integers defined recursively as follows:
\begin{enumerate}
\item if $Q$ selects only a single row $d$ from a base table by specifying a key value $k$, then $\vec{r}(Q):=(r(d))$. (Example: singleton query for Table $T_1$, see Figure \ref{fig:queryRC}) 
\item if $Q$ selects specific rows $d_1, .., d_n$ from a base table, by specifying key values $k_1,..k_n$, then $\vec{r}(Q) = (r(d_1),..,r(d_n))$.
\item if $Q$ selects a single row $d$ from a join of base tables $T_1,..,T_n$ by specifying key values, where $d$ is constructed from rows $d_1$ in $T_1$, $d_2$ in $T_2$ etc., then $\vec{r}(Q)=(r(d_1),..,r(d_n))$.
\item if $Q$ selects some other set of rows, or all rows, from a base table $T$, then $\vec{r}(Q):=(r(T))$
i.e., a vector containing a single integer identifying the most recent change to table $T$. (Example: predicate query for Table $T_2$, see Figure \ref{fig:queryRC}) 
\item if $Q$ is a join, merge, union etc. of queries $Q_1$ and $Q_2$, then $\vec{r}(Q) := \vec{r}(Q_1)\times \vec{r}(Q_2)$ (Example: $Q$ query decomposition by rewrite on Views $V_i$, see Figure \ref{fig:queryRC}) 
\item if $Q$ is the result of aggregation or other SQL operation on previous results from another query $Q'$, then $\vec{r}(Q) = \vec{r}(Q')$, since the data for $Q$ is no fresher than the data in $Q'$.
\end{enumerate}
The specific key values mentioned in 2) above must be explicit at the outermost level of $Q$, and not computed as part of the evaluation. 
This will allow efficient re-computation of $Q$ as there is no need to perform a full evaluation of the query.
Only those rows have to be retrieved that have changed its values since the last result caching. 
Then, a sufficient condition for the results of $Q$ to be unchanged is that $\vec{r}(Q)$ is unchanged. 

We note in passing that the calculation of $\vec{r}(Q)$ for a given query is very efficient. This is obvious because the readCheck vector has only a small dimension because of conditions 2) and 3). If a large set of rows or unknown data sets are selected, the readCheck value of the table is used.   

This definition can be adapted for a multi-database query by creating a string representation combining the name of each contributing database with a string version $s(Q)$ of $\vec{r}(Q)$ (provided that the database agent is required to generate equal strings $s(Q)$ for equal $\vec{r}(Q)$).  
Such a combined string can then be used as an ETag validator for the associated HTTP request as described above.

This readCheck mechanism has been implemented as proof-of-concept using the Pyrrho \cite{Crowe} DBMS. This DBMS is built as a relational database on the .NET framework with pure optimistic concurrency control. It has been rigorously developed to deliver serializable transactions providing full ACID properties and support most of the SQL 2011 syntax.    
Its API allows application data models based on versioned objects. 
Each versioned value contains a readCheck string and updates to versioned objects use PUT, POST and DELETE operations similar to a REST service.

\subsection{Query Scenario with ReadCheck}
\label{querRC}
In order to illustrate the virtual integration and the use of the readCheck mechanism lets assume a distributed query $Q$ issued by $R$ against the Global View $V$. 
\begin{figure}[]
\centering
\includegraphics[width=0.45\textwidth]{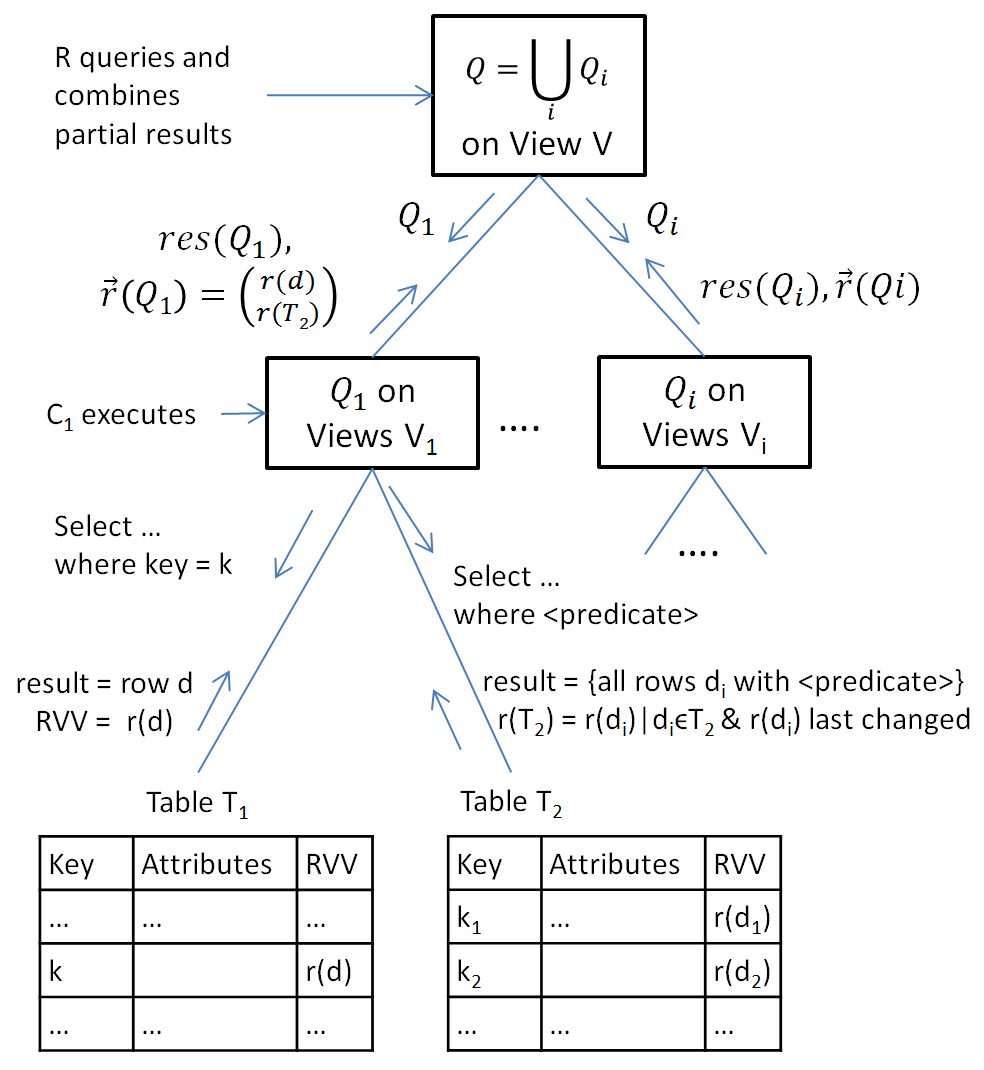}
\caption{A general query scenario collecting data from multiple data providers}
\label{fig:queryRC}
\end{figure}

As in Figure \ref{fig:queryRC} the query $Q$ is rewritten according to the Views $V_i$ and each decomposed query $Q_i$ is executed by the respective contractor $C_i$. It is the contractor's responsibility to provide the readCheck vector for the query result.
 
For example, $C_1$ executes $Q_1$ and computes $\vec{r}(Q_1) = (r(d), r(T))$ where $r(d)$ will be just the RVV value for the selected row $d$ of table $T_1$ and $r(T_2)$ is the RVV value of the row that was last changed in $T_2$. 
The reason for using this value is that the select with predicate could comprise too many rows to include all its RVV efficiently in the readCheck vector. 

The price for this implementation is that it might happen that the predicate query for table $T_2$ is unnecessarily re-executed when the original query $Q$ is issued again. This happens if a row in table $T_2$ has been changed that is not included in the query predicate. 
This "false positive" could be avoided if the readCheck only considers rows meeting the predicate. 
The down side of such an approach would be that the RVV of all rows included in the query must be remembered along with the readCheck value. 
Nevertheless, it is the responsibility of the $C_i$ to determine how to calculate the readCheck validator. 
In fact, the underlying data sources might not be relational nor provide an RVV. So the contractor would need to establish its own readCheck mechanism in accordance with the definition above. 

Depending on the API the results $res(Q_i)$ are delivered to $R$ in a GET-response or as SQL result set. Finally $R$ assembles (e.g., joins, union, etc.) the partial results according to the global view $V$.

\subsection{Global schema design}
\label{ssec:globSchemaD}
Any data warehousing system needs a mechanism for creating a global schema. 
For our experiments we used a schema extension for integrating REST Views into a database schema. 
The BNF-schema definition has therefore been extended to include REST views into a database in the following way: (The \dots represent the former DDL syntax of Pyrrho that is not shown for simplicity. The SQL syntax for Pyrrho can be found in Chapter 7 of \cite{Crowe})

ViewDefinition := \dots $\mid$ CREATE VIEW {[} ViewSpecification{]} id AS GET \{Metadata\} .
 
Metadata := \dots $\mid$ string . 

ViewSpecification := \dots $\mid$ OF '(' TableClause {[}',' TableClause{]}')' {[}UriType{]} . 

UriType := {[}Abbrev\_id{]}'\textasciicircum\textasciicircum'( {[}Namespace\_id{]} ':' id $\mid$ uri ) .
\begin{figure*}[]
\centering
\includegraphics[width=0.95\textwidth]{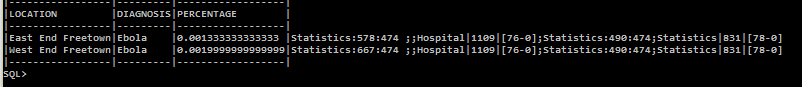}
\caption{Results from the queries of the example scenario}
\label{fig:analysis}
\end{figure*}
We pick up the example from the introduction to illustrate this syntax and implement it as follows:

/* C1: Hospital DB */ \\
\texttt{\footnotesize{ create table D (ID int(11) NOT NULL, name varchar(45), rCode int, birthdate datetime, admission datetime, diagnosis varchar(45), treatment varchar(45), PRIMARY KEY (ID)); \vspace{2mm} \\
 insert into D values \\
(1, 'Joe Soap', 2, date'2003-04-12', date'2014-09-20', 'Ebola', 'IV fluid, electrolytes'), \\
(2, 'Milly Soap', 2, date'2007-10-12', date'2014-10-06', 'Ebola', 'IV fluid, electrolytes'),\\
(3, 'Betty Boop', 1, date'1996-10-12', date'2014-10-06', 'bacterial infection', 'antibiotics'),\\
(4, 'John Bell', 3, date'2009-11-14', date'2014-09-10', 'Ebola', 'electrolytes'),\\
(5,'Benny Hall', 2, date'2007-10-10', date'2014-10-06', 'Ebola', 'IV fluid, electrolytes'); \vspace{2mm} \\
 create view E as select rCode, extract(year from (admission-birthdate)) as age, 
admission, diagnosis, treatment, count(*) as patients
from D group by rCode, age, admission, diagnosis, treatment;}}
 
/* C2: Statistics DB */ \\
\texttt{\footnotesize{ create table H (rCode int NOT NULL, location varchar(45), inhabitants int, under10 int, 10to20 int, 20to30 int,
  over30 int, lastUpdated datetime, PRIMARY KEY (rCode)); \vspace{2mm} \\
  insert into H values \\ 
(1,'Central Freetown',300000, 80000, 75000, 65000, 80000, date'2014-10-20'), \\
(2,'East End Freetown',500000, 150000, 120000, 100000, 130000, date'2014-10-20'),\\
(3,'West End Freetown',200000, 50000, 40000, 40000, 120000, date'2014-10-20');\vspace{2mm} \\
  create view K as select rCode, location, inhabitants, under10, lastUpdated from H; }}
 
/* Requester Schema */ \\
\texttt{\footnotesize{ create view V1 of (rCode int, age int, admissionDate date, diagnosis char, treatment char, patients int) as get 'http://servD1:8180/Hospital/Hospital/E';\vspace{2mm} \\ 
create view V2 of (rCode int, location char, inhabitants int, under10 int, lastUpdated date) as get 'http://servD2:8180/Statistics/Statistics/K';\vspace{2mm}  \\ 
create view V as select * from V1 natural join V2; }}
 
The view definitions of V1, V2 and V here do not copy data from database $D_{11}$ or $D_{21}$; the REST Views V1 and V2 are defined using a URL metadata string for the servers hosting the databases. 
But $R$ can obtain values from C1 and C2 using RESTful operations. 

For instance a) the percentage of young Ebola patients under 10 years of age or b) the total number of treated patients by quarters (of Freetown) and diagnosis can be analysed.
\vspace{2mm} \\
\texttt{\footnotesize{ 
select location, diagnosis, patients/under10)*100 as percentage from V where age $<$ 10; }}

The result of this query is given in Figure \ref{fig:analysis}. The RVV and readCheck information for each row of the result table are provided by the Pyrrho database. Activating the -v flag prints these information on the right of the result table. So the first row of the join has an RVV of "Statistics:578:474". This is the concatenation of the RVV Statistics table H, log position 578 (location "East End Freetown") from transaction 474 and the Hospital contribution to this row (which is blank because of aggregation). 
The result of the above query is produced from the view V which results from two REST GET operations that produce two  ETags and one RVV. ETag "Hospital$|$1109$|$[76-0]" says that the transaction log position was 1109 and any change to table 76 (table D) will invalidate the data. The second REST operation on view K produces RVV "Statistics:490:474" because it returns a single row from the same transaction 474 (location "East End Freetown" ) and the ETag "Statistics$|$831$|$[78-0]" with log position 831 and table 78 (table H).
  
Pyrrho's open-source implementation of the REST view automatically makes V2 an updatable view provided V2 has the necessary permissions on the Statistics DB (owner C2) and generates PUT, POST and DELETE operations on H that result from updates on the view K resp. V2. These operations can be used to curate the data, e.g. with:\\
\texttt{\footnotesize{ update V set inhabitants = 199000, under10 = 49000 where rCode = 3;}} \\ and \\
\texttt{\footnotesize{ delete from V2 where rCode = 5; }}

In real situations, things would not always be so simple, and column renaming and conversion between types and structures would result in a more complex definition of the view V. It seems to us that the REST View concept could be made interoperable, as B does not need to understand the structure or implementation of A's readCheck string, and only requires the property that $s(Q)$ is unchanged if the values read during evaluation of query $Q$ are unchanged.
  
\section{Conclusion and Future Work}
This paper presents the beginnings of a formalism and practical strategy to manage Big Live Data. We show that such a virtual DWH can be based on the REST architecture using a mechanism similar to RFC7232 ETags or row version validators to ensure up-to-date and verifiable results. With the help of readCheck an optimistic concurrency mechanism can be implemented to support distributed transaction processing.
Such a contract would be subject to alteration (e.g., a change in contractual responsibility) and readCheck would help to underpin a secure mechanism for managing such a change as described in \ref{ssec:MgntDistribTa}.

The readCheck mechanism can be implemented efficiently and be further used to optimise query processing so that only changed data sets need to be re-queried and combined with previous results. 
Our pilot implementation of these ideas using the Pyrrho database and complex views shows the applicability of the concept. 

The data transformation from ${V_i}$ to $V$ assumes no data conflict and at present is manually defined but we plan to use metadata to support the data transformation and global schema design. Another point for future work is data curation and making data provenance transparent to the requester of a query.

\end{document}